\documentclass[aps,prb,twocolumn,groupedaddress,showpacs,showkeys]{revtex4-1}

\usepackage{graphicx}
\usepackage{amsmath}
\usepackage{amssymb}
\usepackage{epstopdf}

\begin{document}

\title{Inelastic scattering of xenon atoms by quantized vortices in superfluids}
\author{I.A. Pshenichnyuk$^{1}$}
\author{N.G. Berloff$^{1,2}$
\email[correspondence address: ]{n.g.berloff@damtp.cam.ac.uk}}
\affiliation{$^1$Skolkovo Institute of Science and Technology Novaya St., 100, Skolkovo 143025, Russian Federation}
\affiliation{$^2$Department of Applied Mathematics and Theoretical Physics, University of Cambridge, Cambridge CB3 0WA, United Kingdom }
\date{\today}

\begin{abstract}
We study inelastic interactions of  particles with quantized vortices in superfluids by using  a semi-classical matter wave theory that is analogous to the Landau two-fluid equations, but allows for the vortex dynamics. The research is motivated  by recent experiments on xenon doped helium nanodroplets that show clustering of the impurities along the vortex cores.  We numerically simulate the dynamics of  trapping and interactions of xenon atoms by quantized vortices in superfluid helium and the obtained results can be extended to scattering of other impurities by quantized vortices. Different energies and impact parameters of incident particles are considered. We show that   inelastic scattering  is closely linked to the generation of Kelvin waves along a quantized vortex during the interaction even if there is no capture. The capture criterion of an impurity  is formulated in terms of the binding energy.
\end{abstract}                              


\maketitle

\section{Introduction}

In-depth understanding of the dynamics of quantum fluids, and in particular understanding of processes occurring in quantum turbulence during the formation and evolution of a vortex tangle, requires advanced theoretical modelling and precise experimental probing techniques.
Superfluid helium, being the first quantum fluid available for experiments, and probably the most studied one, generates vorticity at the length scale of angstroms which  makes the direct observation of vortices complicated. 
Indirect measurements usually involve probing vortices  with impurities, which are often used as doping for subsequent optical detection.
Early experiments were performed with electrons \cite{williams-1974, yarmchuk-1979,raymond-1984} and ions \cite{careri-1962, springett-1965}. Later, many types of other  impurities including molecules, molecular clusters and excimers \cite{zmeev-2013} were used as doping to visualize and study quantized vortices.
Modern particle image velocimetry techniques allow to use various kinds of micron size tracer particles to visualize flow patterns in helium \cite{zhang-2004, zhang-2005, bewley-2006}. This methods allow one to trace both the normal and superfluid components (through the interaction with vortices) and thus provide a useful tool to study two-fluid hydrodynamics \cite{poole-2005}.
It is shown experimentally, that the coalescence of metal particles trapped on quantized vortices may lead to the formation of centimetre long wires \cite{moroshkin-2010, gordon-2012}. Such a mechanism provides not only a way to visualize the structure of quantized vortices but also a new approach for producing long metal nanowires.
\citet{zmeev-2013} have shown that a moving vortex tangle can transport molecules through superfluid helium, so the composite particles and molecules can be used to probe the density and orientation of the vortex tangle and  lead to some new and unusual types of matter organization with potentially peculiar properties.

Recently, nanodroplets experiments that embed single atoms and molecules into liquid helium droplets have become a new tool to study  various aspects of superfluid behaviour. 
 In these experiments ultracold helium works as a homogenous matrix for subsequent spectroscopic studies \cite{toennies-1998}. 
 In the experiments of \citet{gomez-2014} femtosecond x-ray coherent diffractive imaging technique was used to demonstrate the existence of vortex arrays in helium droplets through the observation of Bragg patterns. Xenon atoms were used as doping in this experiments. The analysis revealed an unusual form of droplets and line associations of xenon atoms which was explained by the formation of vortices in a rotating helium droplet with subsequent trapping of the xenon atoms at the vortex cores.

Most commonly used   theoretical approach to study the static behavior of impurities in nanodroplets is based on DFT calculations \cite{dalfovo-1995}. It is particularly  successful in finding the minimal energy configurations and so capable of describing  the various aspects of the particle-vortex interaction. The dopant of choice to detect vortices by means of spectroscopic experiments is discussed in \citet{ancilotto-2003} where  the adsorption properties of different atomic impurities are compared. This approach was used to study vortex array equilibrium configurations  in rotating nanodroplets, the  properties of xenon chains trapped by the vortex lines, and to explain shapes and surprising stability of nanodroplets \cite{ancilotto-2014, ancilotto-2015}.
 

Despite the large amount of studies, the details of particle-vortex scattering and especially processes which take place at the vortex core  during the interaction are not well understood firstly, because of the interatomic distances involved, secondly, because there is no first principles models that allow one to describe such a dynamics correctly.
Minimalistic models of particles moving in superfluids at zero temperature usually assume that the Bernoulli's force is a dominant one and that it adequately describes the motion far from the vortex cores \cite{poole-2005}. Close to the vortex, substitution energy based analysis is often used to explain the existence of the potential energy barrier with certain parameters which define the capture and escape probabilities \cite{parks-1966, donnelly-1969}.
At the same time, 3D simulations based on the Gross-Pitaevskii equation \cite{berloff-2000}  and the self-trapping model \cite{gross-1958,clark-1965} demonstrate that the capture of an electron by a quantized vortex is accompanied by the emission of Kelvin waves which propagate along the vortex core and carry a certain portion of energy with them. It makes the particle-vortex scattering process inelastic and renders more detailed energy redistribution analysis. Non-elasticity of the trapping process is similar to inelastic scattering of electrons on molecules, where electrons can be captured by molecules, as a result of internal energy redistribution through the electron-phonon coupling mechanism, forming long-living negative ions.

Xenon particles used as doping in the experiments of \citet{gomez-2014} are very different from electrons, considered in \citet{berloff-2000}. Electron in helium, through its zero-point motion,  forms an electron bubble of a radius of about 16 \AA, that brings about a large (in comparison with the electron) effective mass and the distortion of the soft bubble boundary. This effect for the electrically neutral xenon is minimal and we expect its radius in helium to be of the order of  the size of vortex cores. It results in a significant difference of substitution energies of electrons and xenon atoms. Moreover, atoms are much heavier than electrons and can potentially produce more disturbance  along the vortex lines when  used as doping.

In this paper we develop ideas formulated in \citet{berloff-2000} to study scattering of Xe atoms by quantized vortices in different regimes. We shall elucidate the role of the binding energy and attachment/detachment criteria. The paper is organized as following. We present the model representing the mathematical equivalent of the Landau two-fluid theory which is the basis for our numerical and analytical study in Section II. We discuss motion of a xenon atom next to a straight line quantized vortex and analyze various scenarios of the impurity-vortex interactions in Section III. We conclude with Section IV summarizing the main findings.

\section{Modelling of the vortex-impurity interactions}

\begin{figure}
  \centerline{\includegraphics[width=3.90in]{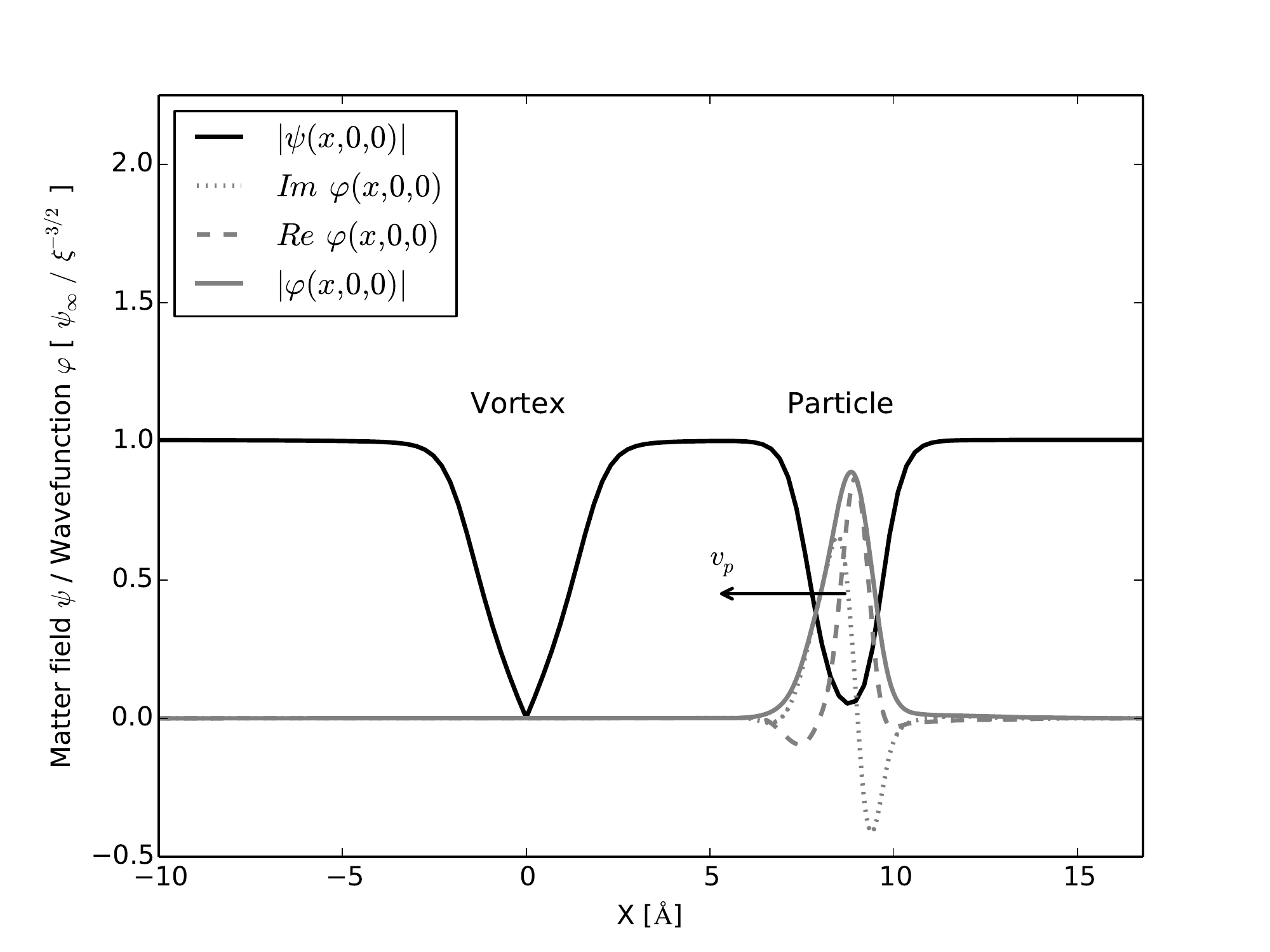}}
  \caption{A snapshot of the one-dimensional  cross-sections of the matter field  $\psi$ and the wavefunction $\varphi$ along the particle-vortex interaction line.  The left drop in the modulus of the amplitude of $\psi$ corresponds to the quantized vortex, the right drop  corresponds to the position of the  xenon atom. The plots of the real and imaginary part of $\varphi$ are given to indicate that the atom is moving towards the vortex.
  \label{selftrapping}}
\end{figure}

\begin{figure*}
  \centerline{\includegraphics[width=\textwidth]{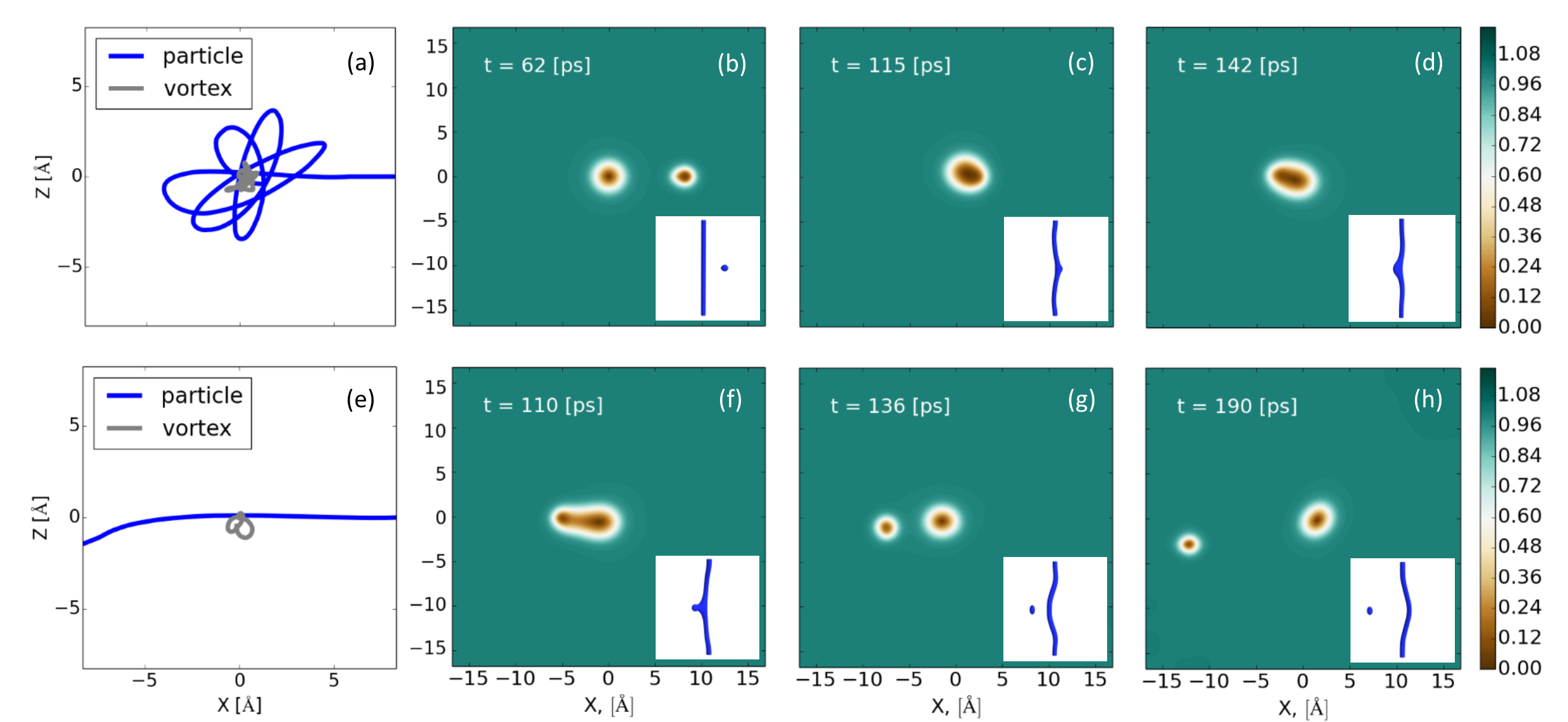}}
  \caption{(Color online) Visualization of two scattering processes with (the top row) and without (the bottom row) trapping. Initially the impurity is located 11 \AA\ away from the vortex and moving with an initial kinetic energy 0.16 meV (the top row) or 0.38 meV (the bottom row). The panels (a) and (e) show the trajectories of the atom (blue line) and the vortex (grey line). Other panels show two-dimensional cross sections of the modulus of the amplitude of the fluid $|{\psi}(x,0,z)|$ at different moments of time. Small three-dimensional insets show the corresponding isosurfaces $|{\psi}(x,y,z)|=0.3{\psi}_\infty$.
  \label{visualization}}
\end{figure*}

A useful approach in modelling the  dynamics and interactions of particles with quantized vortices was originally formulated by Gross  \cite{grant-1974}. In this approach the nonlinear Schr\"odinger equation (NLSE) also known as Gross-Pitaevskii equation (GPE) which describes the wavefunction of a Bose-Einstein condensate is coupled with the  linear Schr\"odinger equation for the particle's wavefunction. In reality  only about 10\% of superfluid helium is in a condensed phase and the fluid is dominated by many-body effects, so its approximation by the condensate order parameter is at best phenomenological. It  was later demonstrated \cite{berloff-2002} that the NLSE in the context of the semi-classical matter field description corresponds to the Landau two-fluids model and, therefore, describes both  the superfluid and the normal fluid as long as the low occupancy modes and their coupling to the highly occupied modes are neglected. The framework of the coupled GP-type equation for the superfluid and normal fluid components  and the equation for the particle's wavefuction can therefore be used at finite temperature. We can further remedy this description and incorporate the equation of state correct for the superfluid helium using a higher order NLSE \cite{berloff-2009,berloff-2014}.   The higher order nonlinearity appears for dense fluids with the equation of state given by a  polynomial expression \cite{pitaevskii}. Such an equation is mathematically equivalent to the Landau two-fluid model and allows one, in addition, to account for the processes associated with quantized vortices. In this sense it provides a framework to describe the behavior of superfluid helium at finite temperatures. In the Appendix A we show how to recover the Landau two-fluid model from our theory.

We formulate the Hamiltonian of the system by introducing various contributions:  the kinetic  and internal energies of superfluid helium $E_{kin}$ and $E_{int}$, the particle-helium interaction  energy $E_h$ (which is the most significant in  the healing layer between the particle and the fluid),  the energy of the xenon particle $E_{p}$ (it includes the kinetic energy of motion $E_{pk}$, which will be discussed later, and the zero-point energy) and explicitly introducing the Lagrange multiplier (the chemical potential) $\mu$  in the view of the constraint on the total number of matter $\int|\psi|^2dV=N$, where $N$ is a number of bosons in the system:
\begin{equation}
  E = E_{kin}+E_{int}+E_h+E_{p}-\mu N,
  \label{full_energy}
\end{equation}
\begin{equation}E_{kin} = \int \frac{\hbar^2}{2m}|\nabla\psi|^2 {dV},\end{equation}
\begin{equation}E_{int} = \int \varepsilon_{int}(|\psi|^2) {dV}, \end{equation}
\begin{equation}E_h = \int U_0|\psi|^2|\varphi|^2 {dV}, \end{equation}
\begin{equation}E_{p} = \int \frac{\hbar^2}{2M}|\nabla\varphi|^2 {dV}. \end{equation}
Here $m$  and $M$ are the masses of the helium atom and the xenon atom, respectively, $\psi$ is classical complex matter field which describes the superfluid and the normal fluid components, $\varphi$ is the wave function of the particle. 
The parameter $U_0=2{\pi}l\hbar^2/M^\ast$ is the local He-Xe interaction potential strength, where $l=3.4$ \AA\ is the He-Xe scattering length \cite{tang-2003}. This value is also close to the sum of the Van der Waals radii of xenon and helium atoms. $M^\ast$ is the reduced mass of the interaction.

The  internal energy functional  is based on the phenomenological equation of state of liquid helium \cite{dalfovo-1995,berloff-2014} and has  the form
\begin{equation} \varepsilon_{int}(n) = -\frac{V_0}{2}n^2-\frac{V_1}{3}n^3+\frac{V_2}{4}n^4, \end{equation}
where $n=|\psi|^2$. 
Coefficients $V_0 = 719\,k_b \,\, K\text{\AA}^3$, $V_1 =3.63\cdot 10^4\,k_b \,\, K\text{\AA}^6$ and $V_2 = 2.48 \cdot 10^6\,k_b \,\, K\text{\AA}^9$ (where $k_b$ is the Boltzmann constant) are chosen to reproduce the binding energy, the density and the sound velocity of liquid helium \cite{pitaevskii}. The Hamiltonian of Eq. (\ref{full_energy}) was used in \citet{berloff-2014} and \citet{pshenichnyuk-2015} to study the multiplication of vortex rings in a superfluid during pressure oscillations.

Performing a variation of the full energy $E$ with respect to $\psi^\ast$ and $\varphi^\ast$  we get the system of equations, where the first one we will refer to as the NLSE-7:
\begin{equation}
  \label{main1}
  \begin{split}
     i\hbar\frac{\partial\psi}{\partial{t}} = -\frac{\hbar^2}{2m} \nabla^2\psi + U_0|\varphi|^2\psi +  \quad\quad\quad\quad\quad\quad\quad \\
    \quad\quad\quad\quad\quad  +(-V_0|\psi|^2-V_1|\psi|^4+V_2|\psi|^6)\psi-\mu\psi, 
  \end{split}
\end{equation}
\begin{equation}
  \label{main2}
  i\hbar\frac{\partial\varphi}{\partial{t}} = -\frac{\hbar^2}{2M} \nabla^2\varphi + U_0|\psi|^2\varphi.
\end{equation}
Function $\varphi$ is normalized  by $\int|\varphi|^2dV=1$. Away from the impurity the fluid wavefunction  acquires its ground state value $\psi=\psi_\infty$  fixing the chemical potential to $\mu=-V_0\psi_{\infty}^2-V_1\psi_{\infty}^4+V_2\psi_{\infty}^6$. For superfluid helium at atmospheric pressure  $\rho_\infty=m\psi_{\infty}^2=145.2$  kg/m$^3$. The healing length, $\xi$,  is given by the characteristic length-scale on which fluid heals itself to the unperturbed value  from  zero value and is determined by matching the kinetic and the potential energy of interactions  $\xi={\hbar}/{\sqrt{2m{\mu}}}=0.92$ \AA. This value also defines the characteristic radius of vortex cores.

We nondimentionalize the system of Eqs. \ref{main1} and \ref{main2} by  $x\to\xi{x}$, $t\to\frac{\xi^2m}{\hbar}t$, $\psi\to\psi_{\infty}\psi$, $\varphi\to\xi^{-3/2}\varphi$ and  numerically integrate it using the 4-th order space discretization and the 4-th order Runge-Kutta time propagation.  Scattering processes are modelled in a computational box of the size $(37.5\xi)^3$ with the resolution of 4 points per healing length $\xi$.
Before the beginning of the dynamical computation, the initial guess for $\psi$ and $\varphi$  is optimized using the imaginary time evolution for a few time steps \cite{berloff-2000}. The initial kinetic energy is given to the particle by multiplying its wavefunction $\varphi$ by the factor $e^{i\mathbf{k}\cdot\mathbf{r}}$.
A typical one-dimensional cross-section of the  fields prepared by this procedure  is shown on Fig. \ref{selftrapping}, where  the initial velocity of the particle points towards the vortex along the plotted axis. The figure shows the fluid and the particle amplitudes and oscillating real and imaginary parts of the particle's wavefunction. Two minima in the  fluid's amplitude  correspond to the vortex and the depletion due to the repulsive interactions with the impurity.

For comparison, we have also  performed computations using a simple classical model for the interaction of the particle with a vortex. It is based on the theory developed in \citet{poole-2005} and \citet{sergeev-2006} to study the motion of tracer particles in superfluid helium in the  presence of  quantized vortex lines. This approach takes into account a number of forces which are associated with both superfluid and normal components of superfluid helium. At sufficiently low temperatures (below 1 K) where the superfluid component dominates, this approach reduces to the Newton equation of motion for the xenon atom with the dominating effect coming from   the Bernoulli's  force that  appears as the result of the existence of the pressure gradients, produced by the inhomogenious velocity field of the vortex. The equation of motion reads
\begin{equation}
M\frac{d\mathbf{v}_p}{dt} = \int\limits_S P(\mathbf{r}) \hat{\mathbf{n}} dS,
\end{equation}
where $P(\mathbf{r})$ is the superfluid pressure field, $\hat{\mathbf{n}}$ is the unit vector normal to the surface of the particle $S$  and the integral is taken over the impurity's surface which  is assumed to be spherical with the  radius $2.4\,$\AA. This value is close to the Van der Waals radii of xenon atom and is consistent with the scattering length used in the NLSE-7 modeling. Being based on the classical Euler equations  this theory can't handle properly the capture of particles by vortex lines \cite{poole-2005} as it can  describe only  elastic scattering and  the particle's motion away from any vortex cores.

\section{Interactions of the vortex with a moving impurity}

\begin{figure}
  \centerline{\includegraphics[width=3.90in]{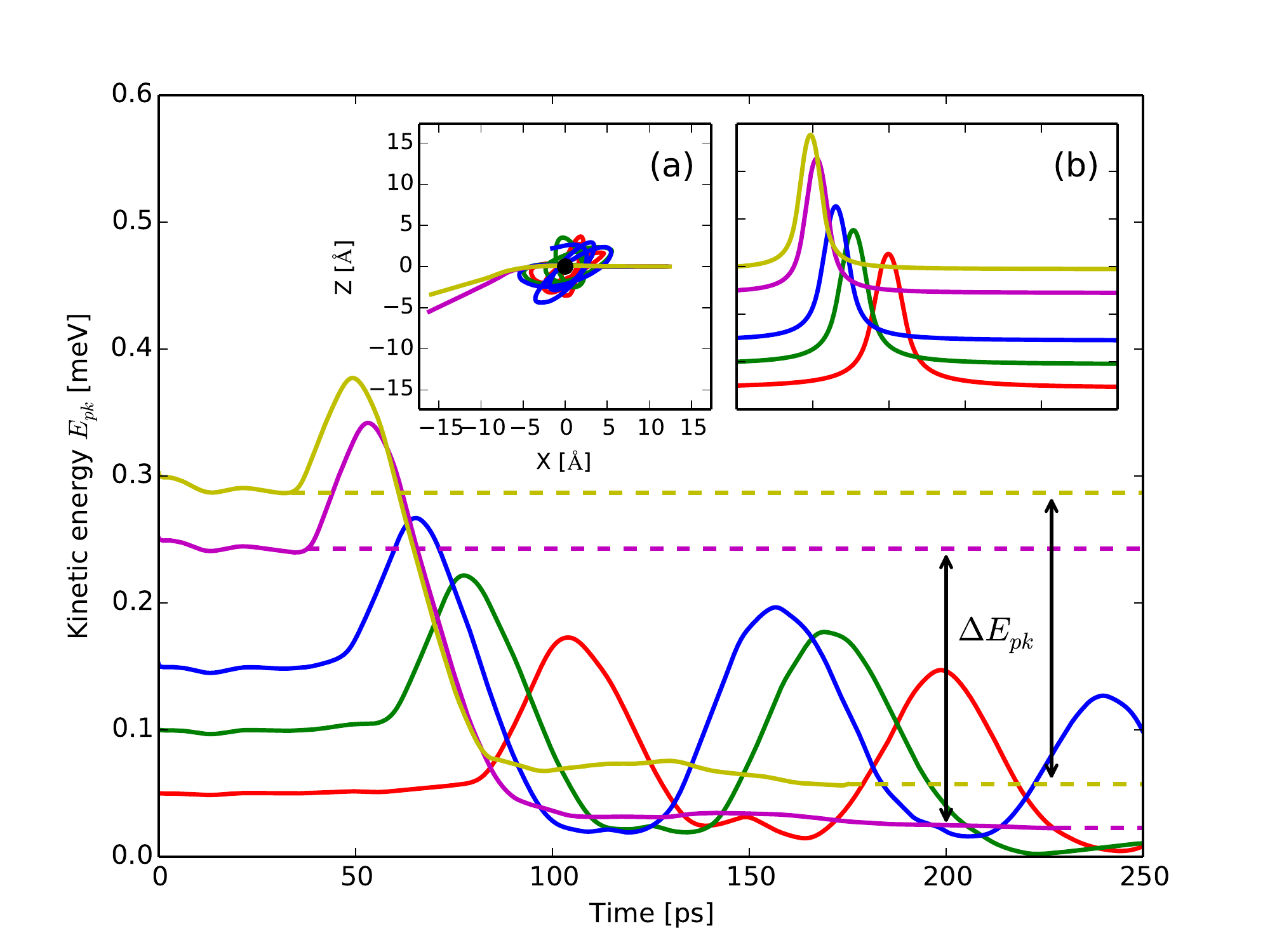}}
  \caption{(Color online) Time evolution of kinetic energies of particles during the scattering events for different initial velocities. The insets show (a)  the corresponding trajectories of the impurity and (b) the corresponding time evolutions of the kinetic energy based on the classical Bernoulli's force  calculations. Curves in the inset (b) are plotted in the same time/energy window as the main figure. 
  \label{scattering_kinetic}}
\end{figure}

\begin{figure}
  \centerline{\includegraphics[width=3.90in]{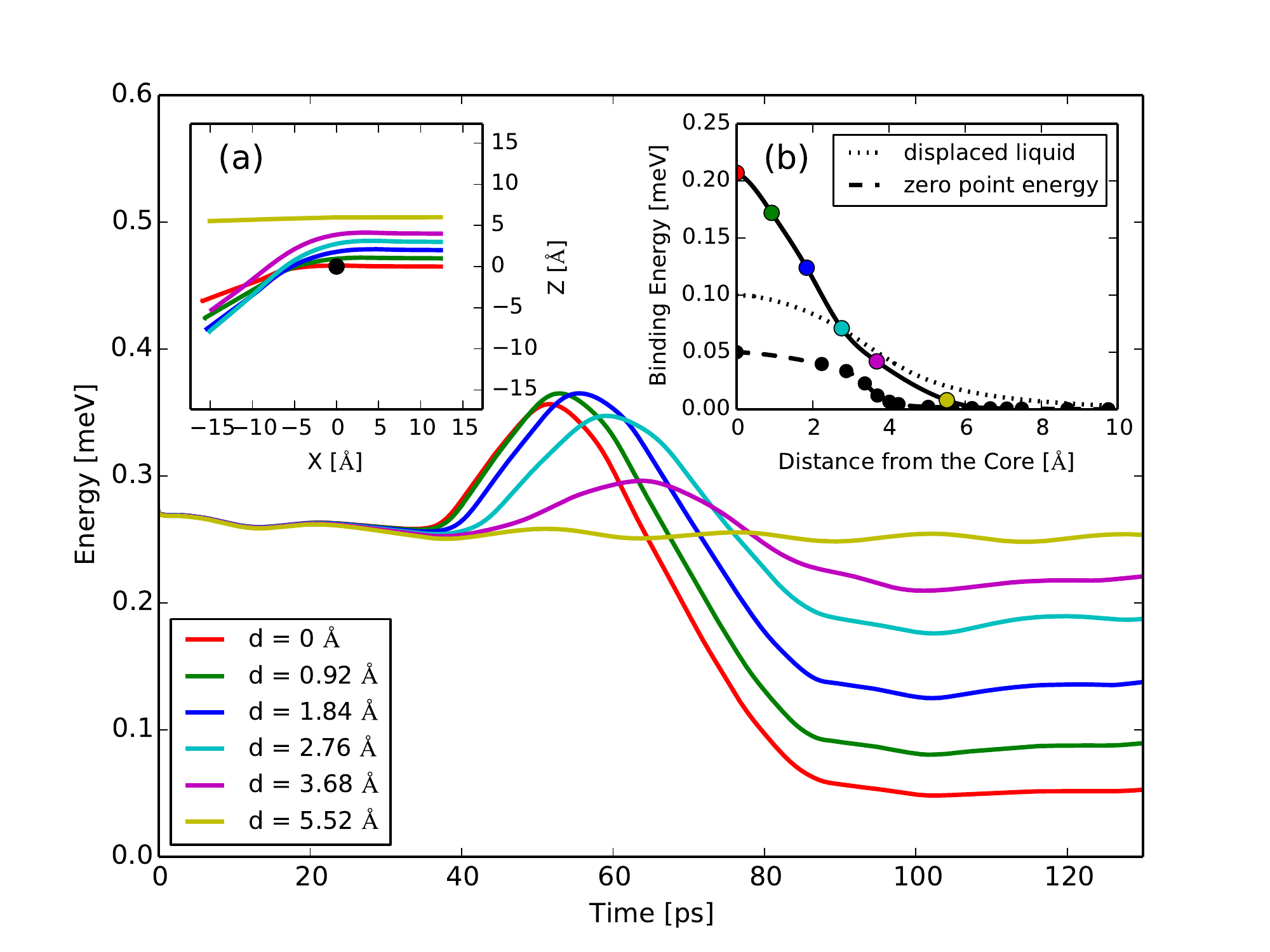}}
  \caption{(Color online) Kinetic energies of particles scattered with different values of the impact parameter $d$. The insets show  (a) the corresponding trajectories of the impurity and of the vortex core and (b) the inelastic energy loss $\Delta{E}$  for different $d$. Numerical estimations of the two main contributions to $\Delta{E}$ are shown with dashed and dotted lines in (b).
  \label{scattering_impact}}
\end{figure}

\begin{table*}
  \caption{Binding energies $\Delta{E}_0$ of the xenon atom and the electron  attached to the quantized vortex. Values are obtained using stationary numerical computations. Corresponding energy terms and their contributions to the total binding energy are shown. \label{binding_energy}}
  \begin{ruledtabular}
  \begin{tabular}{cccccc} 
                  & $\Delta{E_0}$, meV & $\Delta{E_{kin}}$, meV & $\Delta{E_{int}}$, meV & $\Delta{E_{h}}$, meV & $\Delta{E_{p}}$, meV \\
    \colrule
    Xenon    & 0.19 & 0.10 (53\%) & 0.01 (5\%)& 0.03 (16\%) & 0.05 (26\%) \\    
    Electron & 6.46 & 5.48 (85\%) & 0.53 (8\%)& 0.03 ($<$0.5\%) & 0.42 (7\%) \\
  \end{tabular}
  \end{ruledtabular}
\end{table*}

It is energetically favourable for a particle to be captured by a vortex \cite{dalfovo-2000,berloff-2000} since the particle-vortex binding energy, $\Delta{E}_0 $, defined as the difference between the energy of the system when the vortex and the particle are far away from each other and the energy of the particle located on the vortex core, is positive.  Both energies have the same logarithmic divergencies linked to the divergence of the energies of the vortex velocity field which falls as $\hbar/mr$ with the distance $r$ away from the vortex. The standard approach to deal with such integrals is to introduce a finite radius, $R$, of integration, which gives  the energy of the vortex line in the NLSE-7 to be (see Appendix B for derivation)
\begin{equation}
E_{\text{vort}} = L\pi{\psi^2_{\infty}}\frac{\hbar^2}{m}\ln \left( \frac{1.39R}{\xi} \right),
\end{equation}
where $L$ is the length of the straight line vortex.
 The logarithmic divergencies of the vortex-particle complex and the vortex line cancel out to give proper integrals that are evaluated numerically to give   $\Delta{E}_0 = 0.19 \,\,\text{meV}$ for the xenon atom  and $\Delta{E}_0 = 6.46 \,\,\text{meV}$ for the electron bubble, see Table \ref{binding_energy}. We have also considered various  energy contributions to the binding energy to show that the main contribution comes from the kinetic energy: when trapped the impurity replaces a significant volume of circulating fluid \cite{parks-1966,berloff-2000}. The second contribution to $\Delta{E}_0$ comes from the zero-point energy of the particle ${E}_\text{p}$ (the particle doesn't move and there is no kinetic energy component in ${E}_\text{p}$). It is connected with the confinement radius of its wave function and the uncertainty principle. Since the vortex core is "hollow" inside it provides a weaker confinement than the bulk helium, decreasing the uncertainty in momentum and the zero-point energy. The density of the xenon atom captured by the vortex has an ellipsoidal shape, in contrast with  the spherical shape in the bulk. Changing of the form and staying inside the vortex core rearranges the healing layer between the particle and the fluid, which decreases the healing energy $E_{h}$ as well. The internal energy change $\Delta{E_{int}}$ is negligible.

When the xenon atom approaches the vortex core and gets trapped it releases a portion of energy $\Delta{E}_0$. In comparison with a weakly interacting condensate modelled by the GPE a system described by the NLSE-7 is not as compressible, so only a negligible amount of energy is  converted into sound waves   \cite{berloff-2004}. The dominant effect is the generation of  the Kelvin waves  along the vortex line, carrying the excess energy away from the interaction site \cite{berloff-2000}.
The emission of the Kelvin waves plays an important role during the scattering of xenon atoms on vortices, when particles possess some initial kinetic energy. If the particle's kinetic energy is large enough the particle may pass through the vortex. Since some portion of the full energy stay locked in the Kelvin waves, the impurity should sacrifice the same amount of its kinetic energy, and  slow down or get trapped. This makes the particle-vortex scattering a purely inelastic process. It has a certain resemblance to the well-studied inelastic scattering of electrons on molecules, where vibrational modes of the molecule may accept a certain portion of energy, keeping the electron trapped for a long time \cite{pshenichnyuksa-2010, pshenichnyuksa-2011, pshenichnyuksa-2014, pshenichnyuksa-2006}. 
The difference between our case and the scattering of electrons on molecules is that the spectrum of the Kelvin waves is continuous \cite{barenghi-1985, baggaley-2011d} (while molecular electronic and vibrational spectra are discrete) and the particle-vortex interaction is likely to be non-resonant. A discrete spectrum can be introduced in our system by considering a narrow channel where  the vortex line is pinned by the container walls and therefore only certain wavelengths of the Kelvin waves can be exited.

First we consider a head-on collision of the impurity with the vortex line. On Fig. \ref{visualization} we present the visualization of two scattering processes with (the top row) and without (the bottom  row) trapping. The vortex line is initially located  along the vertical axis. The particle is placed $11\,\text{\AA}$ away from the vortex, with the initial velocity directed towards  the vortex. The left panels in each row show the trajectory of the particle (blue line), recorded at the position of particle's density maximum. Motion of a selected point of the vortex core (slightly above the particle) is shown by the grey line. The maximum amplitude of the Kelvin waves generated is approximately $1\,\text{\AA}$ in this case. The Kelvin waves appear in both cases whether or not the trapping took place. If the particle detaches from the vortex core  we also detect its vibrational motion \cite{gross-1968}.  Other panels  on Fig. \ref{visualization} illustrate the dynamics of the vortex interactions with the impurity via the time snapshots of the absolute value of the matter field $|\psi|$. 

On Fig.\ref{scattering_kinetic} the kinetic energy of the particle\cite{svistunov}
\begin{equation}
E_{pk} = \frac{\hbar^2}{2M} \left[ \text{Im} \int \varphi^{\ast} \nabla \varphi dV \right]^2
\end{equation}
is shown as the function of time for the different initial velocities of the impurity. On Fig. \ref{scattering_kinetic}(a) we present corresponding trajectories of the impurity. The initial position of the vortex is shown with a black dot. In three cases out of five, which correspond to lower initial energies, the xenon atom gets trapped. Figure \ref{scattering_kinetic}(b)  shows the results obtained for the same initial configurations using the Bernoulli force based classical approach as described in the previous section.  Such a minimalistic model draws a purely elastic scattering picture in a centrally symmetric potential. The Bernoulli's force causes particle to accelerate when it approaches the vortex and to decelerate when it moves away from it. The width and position of the resulting peak depends on the initial velocity. There are obvious similarities with NLSE-7 results with respect to the positions of peaks which indicates that the Bernoulli force accurately describes the dynamics of particles outside of the interaction region (where the separation between the impurity and the vortex  is larger than 5\AA, according to our simulations). The height of the peaks is higher in classical computations, as the energy in our model is being continuously redistributed between various terms.
The trapped particles oscillate around  the vortex core along elliptic trajectories with an amplitude of about $5$ \AA. Their kinetic energy time dependence contains multiple maxima as it is shown on Fig.\ref{scattering_kinetic}. During such motion the particle's energy continuously dissipates and the amplitude of peaks goes down in time. It is accompanied by the increase in the healing energy while no further increase in the Kelvin waves amplitude is detected.  The time evolution for trapped particles is computed for $1$ ns to ensure that the  particle does not detach.

 When the particle does not become trapped there is an energy drop $\Delta{E}_{pk}{\approx}0.2$ meV, given by the difference of the initial and final kinetic energies, characterizing the non-elasticity of the process. The value of $\Delta{E}_{pk}$ within the accuracy of the simulation coincides with the binding energy $\Delta{E}_0$ which shows  that the portion of energy equal to $\Delta{E}_0$ is being transferred to the Kelvin waves  during the interaction causing the drop in  the kinetic energy of the particle. If the xenon initial energy is lower than $\Delta{E}_0$ it can not escape and gets trapped by the vortex line. This value defines the capture criteria for xenon atoms by vortices in superfluid helium at  low temperatures.

In some regimes we observe the splitting of  the particle wave function, $\varphi$, between two spacial locations. During the detachment, small part of the particle wave function may remain attached to the vortex. This reflects the probabilistic nature of the process, and is interpreted as the existence of some finite probability of the particle to get captured even at high energies. In cases which are characterized as scattering regimes with no trapping this probability is usually less than 5 percent (defined as the portion of the trapped mass of  the particle). We stress that despite the fact that the superfluid is modelled in terms of classical fields, for the particle we have usual linear Shr\"odinger equation, which describes quantum effects typical for a particle in a potential well. 
Nevertheless,  the interaction picture in our models is more classical than quantum, there exists a sharp border between attachment and detachment regimes.

Next we consider the scattering and trapping of  the xenon atom which is off-set from a vortex line in the direction of its motion. As was shown for the head-on collision, the main contribution to the binding energy comes from  $\Delta{E}_\text{kin}$ (see Table \ref{binding_energy}), which represents the kinetic energy of superfluid displaced from the vortex velocity  field by the particle. The value of $\Delta{E}_\text{kin}$ is expected to be smaller than the one in Table \ref{binding_energy} if the particle is placed at a certain distance from the vortex core, since the superfluid velocity decreases with this distance. It should be reflected in scattering events when the particle passes at a certain distance from the core. On Fig. \ref{scattering_impact} we present results for different values of an impact parameter $d$ (the minimal distance between the straight line trajectory  of the particle in the absence of the vortex and position of the vortex core). The particle trajectories are plotted on Fig. \ref{scattering_impact}(a). It is clearly seen how the inelastic energy drop, $\Delta{E}_{pk},$ decreases with $d$. This dependence is plotted on Fig. \ref{scattering_impact}(b) by the solid line. We have shown above that for the head-on collision $\Delta{E}_{pk}$  coincides with the binding energy $\Delta{E}_0$. It is impossible to use the same method of evaluation for $\Delta{E}_0$ when $d{\neq}0$, since such configurations are not steady. In Fig. \ref{scattering_impact} (b) we show rough numerical estimations how $\Delta{E}_\text{kin}$ (dashed line) and ${E}_\text{p}$ (dotted line) depend on the distance from the core.  ${E}_\text{p}$ is associated with the zero-point energy variation during the interaction, and not with the kinetic energy of the particle. Their sum constitute almost 80\% of $\Delta{E}$. This analysis  again points out that  the effective radius of interaction for xenon atoms and quantized vortices in helium is about $5\,\text{\AA}$ (see Fig. \ref{scattering_impact}(b)).

The theory described in this manuscript can be easily extrapolated to other types of particles.  To illustrate, in Table \ref{binding_energy} we compare binding energies $\Delta{E_0}$ with corresponding components for the xenon and an electron. The fraction of $\Delta{E}_\text{kin}$ in the binding energy is much lager for the electron than for the xenon in the view of the large radius of the electron bubble as compared to the xenon radius and therefore larger volume of displaced fluid. The value of $\Delta{E_{kin}}$ obtained here for the electron is close to the one obtained using the GPE \cite{berloff-2000}. 
For the basic analysis of the electron capture we may assume $\Delta{E_0}\approx\Delta{E_{kin}}$ and compute it using the model suggested by \citet{parks-1966}. 

\section*{Conclusion}

In this manuscript we studied the inelastic scattering of xenon atoms on quantized vortices in liquid helium. The theoretical framework based on the modified version of the self-trapping wave function approach is used to model the dynamics of the vortex-particle interactions. It is argued that NLSE-7  as a model of superfluid helium is mathematically analogous to the Landau two-fluid model and in this sense can be used to model the dynamical effects in superfluid helium.  It is shown that Kelvin waves are excited along the vortex filament during the interaction with a particle whether or not the particle is trapped at the vortex core, keeping a certain portion of energy and providing a mechanism for the inelastic trapping or scattering of particles. The simple capture criteria for xenon atoms is formulated. It states that in head-on collisions the particle is captured if its kinetic energy is less than the binding energy, which is equal to 0.2 meV for xenon. For the nonzero impact parameter $d$ the capture criteria becomes weaker and starting from $d{\approx}6\,\,\text{\AA}$ practically no capture occurs.

\appendix
\section{Derivation of the Landau two-fluid model from classical field equations}

The idea to use classical fields approximation to model superfluid helium can be traced back to the works of Putterman and Roberts \cite{putterman-1983}. Using the scale separation in GPE they derived an equivalent set of kinetic equations which describe both the condensate and the thermal cloud, as well as their interaction, so the classical field $\psi$ is no longer directly associated with the condensate. Instead, the  separation of scales  leads to association of  the slowly varying, large-scale, background field with the superfluid component, and the short, rapidly evolving excitations with the normal component.  Therefore, $\psi$ in this context gives rise to both components. This result allows one to generalize the classical field approach and perform finite temperature GPE based computations \cite{berloff-2002, berloff-2007}. Another important step in this direction was made by demonstrating the equivalence of GPE and the Landau two-fluid model using the local gauge transformation \cite{coste-1998, geurst-1980,salman-2011}. Gauge field in this case is related to additional macroscopic degrees of freedom and allows one to switch from one-fluid to two-fluid system. In this section we use the similar procedure to demonstrate the equivalence of NLSE-7 and Landau two-fluid model. 

The Lagrangian density for NLSE-7 reads
\begin{equation}
  \begin{split}
\mathcal{L}_0 = \frac{i\hbar}{2} \left[ \psi\dot{\psi}^{\ast}-\psi^{\ast}\dot{\psi} \right] 
+ \frac{\hbar^2}{2m}|\nabla{\psi}|^2 \\
- \frac{V_0}{2}|\psi|^4 - \frac{V_1}{3}|\psi|^6 + \frac{V_2}{4}|\psi|^8.
  \end{split}
\end{equation}
We apply the local gauge transformation $\psi \rightarrow \psi e^{i\alpha(\mathbf{r},t)m/\hbar}$, which provides 4 additional independent variables for the nonzero temperature two-fluid model description.
Newly introduced scalar and vector fields are denoted as $\xi \equiv \dot{\alpha}(\mathbf{r},t)$, $\mathbf{A}\equiv - \nabla\alpha(\mathbf{r},t)$. They appear as additional terms in the Lagrangian 
\begin{equation}
\begin{split}
  \mathcal{L}_1 = \mathcal{L}_0 + m\xi |\psi|^2 + \frac{m}{2}A^2|\psi|^2 \\
  - \frac{\hbar}{2i}  \mathbf{A} \cdot \left[ \psi^{\ast}\nabla\psi - \psi\nabla\psi^{\ast} \right]
\end{split}
\end{equation}
Switching to hydrodynamic variables $\rho$ and $\phi$ such that
\begin{equation}
  \psi = \sqrt{\frac{\rho(\mathbf{r},t)}{m}} e^{i\phi(\mathbf{r},t)m/\hbar},
\end{equation}
we get
\begin{equation}
\begin{split}
  \mathcal{L}_0 = \rho\dot{\phi} + \frac{\hbar^2}{8m^2\rho}(\nabla\rho)^2 + \frac{\rho}{2}   (\nabla\phi)^2 \\
 +\left\{ -\frac{V_0}{2} \frac{\rho^2}{m^2} -\frac{V_1}{3}\frac{\rho^3}{m^3}+\frac{V_2}{4}\frac{\rho^4}{m^4}\right\},
\end{split}
\end{equation}
\begin{equation}
  \mathcal{L}_1 = \mathcal{L}_0 + \frac{A^2\rho}{2} + \rho\xi - \rho \mathbf{A}\cdot \nabla\phi .
\end{equation}
According to \citet{coste-1998} we link scalar and vector fields with physical variables in a following way
\begin{equation}
  \xi = \eta(\rho,s) + \mathbf{v}_n\cdot\mathbf{A},
\end{equation}
\begin{equation}
  \mathbf{A}=\chi(\rho,s)(\nabla\phi - \mathbf{v}_n),
\end{equation}
where $\chi$ and $\eta$ are Galilean invariant scalars which are functions of density and entropy only. Thus, the new variables which we add to the model are the normal fluid velocity $\mathbf{v}_n$ and the entropy $s$.
The Lagrangian reads (curly brackets are used to highlight the nonlinear part of NLSE-7)
\begin{equation}
\begin{split}
  \mathcal{L}_1 = \rho\dot{\phi} + \frac{\rho}{2}(\nabla\phi)^2 + \frac{\hbar^2}{8m^2\rho}(\nabla\rho)^2 \quad \quad \quad \quad \\
+\left\{ -\frac{V_0}{2} \frac{\rho^2}{m^2} -\frac{V_1}{3}\frac{\rho^3}{m^3}+\frac{V_2}{4}\frac{\rho^4}{m^4} \right\}
 +\rho\eta+\rho\chi\mathbf{v}_n \cdot (\nabla\phi-\mathbf{v}_n) \\
+ \frac{\rho\chi}{2}(\chi-2)(\nabla\phi)^2 + \rho\chi(1-\chi)\nabla\phi\cdot\mathbf{v}_n + \frac{\rho}{2}\chi^2v_n^2  \quad \quad
\end{split}
\end{equation}
The Euler-Lagrange equation for $\phi$ is
\begin{equation}
\frac{\partial\mathcal{L}_1}{\partial\phi} - \nabla\frac{\partial\mathcal{L}_1}{\partial(\nabla\phi)}
-\frac{\partial}{\partial{t}}\frac{\partial\mathcal{L}_1}{\partial\dot{\phi}}=0.
\end{equation}
Substituting $\mathcal{L}_1$ and computing derivatives we get 
\begin{equation}
\frac{\partial\rho}{\partial{t}} + \nabla\cdot \left[ \mathbf{v}_n\rho\chi(2-\chi) +\nabla\phi\rho(1-\chi)^2\right] = 0.
\end{equation}
Recalling that $\mathbf{v}_s = \nabla\phi$ and introducing notations $\rho(1-\chi)^2=\rho_s$ and $ \rho\chi(2-\chi)=\rho_n$ we obtain the first equation of Landau's model (the equation for mass conservation).

The second Landau equation (the equation for the superfluid velocity) is derived from the Euler-Lagrange equation for $\rho$ (one should recall that both $\chi$ and $\xi$ are functions of $\rho$)
\begin{equation}
\frac{\partial\phi}{\partial{t}}+\frac{1}{2}(\nabla\phi)^2+\tilde{\mu} = \frac{\hbar^2}{2m^2}
\left[ \frac{(\nabla\rho)^2}{4\rho^2} + \frac{\nabla^2\rho}{2\rho}\right],
\end{equation}
where
\begin{equation}
\begin{split}
\tilde{\mu} \equiv \eta+\rho\frac{\partial\eta}{\partial\rho} + 
\left\{ -\frac{V_0}{m^2}\rho -\frac{V_1}{m^3}\rho^2 + \frac{V_2}{m^4}\rho^3 \right\} \\
-\frac{1}{2}\left[ 2\rho(1-\chi)\frac{\partial\chi}{\partial\rho} + \chi (2-\chi) \right] (\mathbf{v}_n - \mathbf{v}_s)^2.
\end{split}
\end{equation}
The difference of this result with the one obtained in \citet{salman-2011} is contained in $\tilde{\mu}$. Polynomial function of $\rho$ in curly brackets appears instead of single linear term in GPE. This doesn't change the main logic of the original derivation.

Remaining two equations of the two-fluid model should be derived from additional constraints which appear as Lagrange multipliers in $\mathcal{L}_1$ and correspond to  the conservation of entropy and relative fluid velocity. This part of the derivation is the same for GPE and NLSE-7 \cite{coste-1998}.

\section{Energy of the vortex in NLSE-7}

\begin{figure}
  \centerline{\includegraphics[width=3.90in]{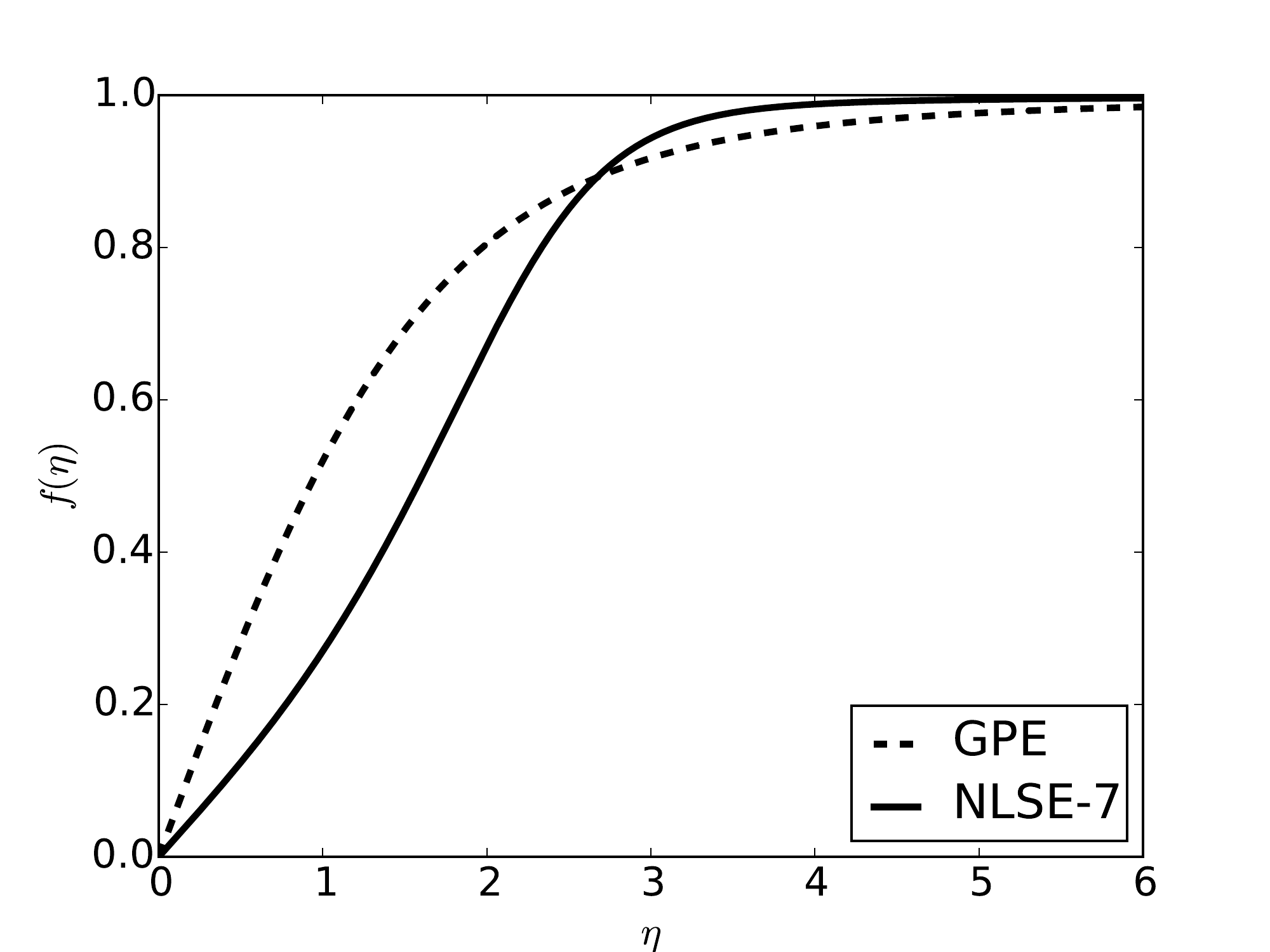}}
  \caption{Dimensionless radial part $f$ of vortical solutions in NLSE-7 and GPE models as a function of dimensionless coordinate $\eta$. The superfluid density is given by $n=f^2\psi_{\infty}^2$
  \label{core_structure}}
\end{figure}

Stationary NLSE-7 reads
\begin{equation}
  \begin{split}
     -\frac{\hbar^2}{2m} \nabla^2\psi  +(-V_0|\psi|^2-V_1|\psi|^4+V_2|\psi|^6)\psi-\mu\psi = 0.
  \end{split}
\end{equation}
We switch to cylindrical coordinates ($r$,$\Phi$,$z$) and search the vortical solution in the form\cite{pitaevskii}
\begin{equation}
\psi = e^{i\Phi}|\psi_0(r)|.
\label{vortical_solution}
\end{equation}
Using the dimensionless units such that $|\psi_0|=f(\eta)\psi_{\infty}$ and $r = \eta \xi$ along with definitions of chemical potential $\mu$ and healing length $\xi$ given in the section II we can derive the following equation for the radial part $f$ of the vortical solution of Eq. (\ref{vortical_solution})
\begin{equation}
\frac{1}{\eta}\frac{d}{d\eta}\left( \eta \frac{df}{d\eta} \right) - \left(\frac{1}{\eta^2} + 1\right) f + c_1f^3+c_2f^5-c_3f^7 = 0,
\end{equation}
where $c_1=2.19329$, $c_2=2.42001$ and $c_3=3.61330$. To obtain the details of the vortex core structure this equation is solved numerically using the shooting method with initial conditions $f(0)=0$, $\frac{df(0)}{d\eta}=k$. Parameter $k$ is chosen to fulfil another known physical boundary condition $f(\infty)=1$. The resulting function is plotted on Fig. \ref{core_structure} along with the vortex amplitude of  the GPE for comparison.

The energy of the vortex is given by the full Hamiltonian of the system, where $\psi$ represents the vortex solution computed above
\begin{equation}
\begin{split}
  E_v = \int \left( \frac{\hbar^2}{2m}|\nabla{\psi}|^2  -\frac{V_0}{2}|\psi|^4 - \frac{V_1}{3}|\psi|^6 + \right. \\
 \left. + \frac{V_2}{4}|\psi|^8   - {\mu}|\psi|^2 \right) dV -E_{gs}.
\end{split}
\end{equation}
The ground state energy $E_{gs}$  is given by the Eq. (\ref{full_energy}) with $\psi=\psi_{\infty}$. Substituting  the solution of Eq. (\ref{vortical_solution}) and using dimensionless variables as above we can express this integral in terms of $f$
\begin{equation}
\begin{split}
  E_v = \frac{{\pi}L{\psi^2_{\infty}}\hbar^2}{m} \int\limits_0^{R/\xi} \left\{ \left(\frac{df}{d\eta}\right)^2 + \left( \frac{1}{\eta^2}+ 1 \right) f^2   \right. \\
 \left. -\frac{c_1}{2}f^4-\frac{c_2}{3}f^6+\frac{c_3}{4}f^8  -c_4 \right\} \eta d\eta,
\end{split}
\end{equation}
where $c_4 = 0.00001$. This formula gives the energy of the vortex enclosed in a cylindrical volume of length $L$ and radius $R$.

If we consider large $R\gg\xi$ this formula can be significantly simplified, since $f\to{1}$ fast with $R$. We simply consider $f=1$ when $R>a$, where $a$ is some constant. The integral splits into two parts and the second one can be taken analytically. The resulting formula reads
\begin{equation}
E_v = \frac{{\pi}L{\psi^2_{\infty}}\hbar^2}{m} \ln{\left(\frac{1.39R}{\xi}\right)}.
\end{equation}
The numerical coefficient $1.39$ obtained here differs from the coefficient in the similar GPE formula which is equal to $1.46$\cite{pitaevskii}.

\bibliography{paper}

\end{document}